\documentclass[longbibliography,twocolumn,superscriptaddress,showpacs,tightenlines]{revtex4-1}

\usepackage{amsmath,amsfonts,amssymb}
\usepackage{xcolor}
\usepackage{graphicx}
\usepackage{microtype}
\usepackage{bm}
\usepackage[normalem]{ulem}

\usepackage[T1]{fontenc}
\usepackage{hyperref}

\newcommand{\bew}{\begin{widetext}}
\newcommand{\ew}{\end{widetext}}

\newcommand{\bp}{\mathbf{p}}
\newcommand{\bq}{\mathbf{q}}
\newcommand{\bv}{\mathbf{v}}

\newcommand{\br}{\mathbf{r}}

\newcommand{\bff}{\mathbf{f}}

\newcommand{\bn}{\mathbf{n}}

\newcommand{\bh}{\mathbf{h}}
\newcommand{\bfm}{\mathbf{m}}

\newcommand{\bsf}[1]{\textsf{\textbf{#1}}}

\newcommand{\beq}{\begin{equation}}
	\newcommand{\eeq}{\end{equation}}
\newcommand{\beqn}{\begin{eqnarray}}
	\newcommand{\eeqn}{\end{eqnarray}}

\DeclareMathOperator{\Tr}{Tr}
\newcommand{\Q}{\bsf{Q}}
\newcommand{\Hij}{\bsf{H}}
\newcommand{\Om}{\mathbf{\Omega}}
\newcommand{\E}{\bsf{E}}
\newcommand{\M}{\bsf{M}}
\usepackage{subfigure}

\newcommand{\sbell}[1]{\textcolor{teal}{#1}}
\newcommand{\AM}[1]{\textcolor{red}{#1}}

\usepackage{soul}

\begin{document}
\title{Ordering, spontaneous flows and aging in active fluids  depositing tracks}
\author{Samuel Bell}
\email{samuel.bell@sorbonne-universite.fr}
\affiliation{Sorbonne Université, CNRS, Institut de Biologie Paris-Seine (IBPS), Laboratoire Jean Perrin (LJP), F-75005, Paris}

\author{Joseph Ackermann}
\email{joseph.ackermann@sorbonne-universite.fr}
\affiliation{Sorbonne Université, CNRS, Institut de Biologie Paris-Seine (IBPS), Laboratoire Jean Perrin (LJP), F-75005, Paris}

 \author{Ananyo Maitra}
 \email{nyomaitra07@gmail.com}
 \affiliation{Sorbonne Université, CNRS, Institut de Biologie Paris-Seine (IBPS), Laboratoire Jean Perrin (LJP), F-75005, Paris}
\affiliation{LPTM, CNRS/CY Cergy Paris Université, 
F-95032  Cergy-Pontoise cedex, France}

\author{Raphael Voituriez}
\email{raphael.voituriez@sorbonne-universite.fr}
\affiliation{Sorbonne Université, CNRS, Institut de Biologie Paris-Seine (IBPS), Laboratoire Jean Perrin (LJP), F-75005, Paris}

\begin{abstract}
Growing experimental evidence shows that cell monolayers can induce long-lived perturbations to their environment, akin to footprints, which in turn influence the global dynamics of the system. Inspired by these observations, we propose a comprehensive theoretical framework to describe systems where an active field dynamically interacts with a non-advected footprint field, deposited by the active field. We derive the corresponding general hydrodynamics for both polar and nematic fields. Our findings reveal that the dynamic coupling to a footprint field induces remarkable effects absent in classical active hydrodynamics, such as  symmetry-dependent modifications to the isotropic-ordered transition, which may manifest as either second-order or first-order,  alterations in spontaneous flow transitions, potentially resulting in oscillating flows and rotating fields, and initial condition-dependent aging dynamics characterized by long-lived transient states. Our results suggest that footprint deposition could be a key mechanism determining the dynamical phases of cellular systems, or more generally active systems inducing long-lived perturbations to their environment.
\end{abstract}
\maketitle

Active matter theory, and in particular active hydrodynamics \cite{Toner:2005,marchetti2013hydrodynamics,Julicher:2018aa}, have proved over the last years to be a powerful tool to describe the dynamics of living systems at various scales, from biopolymer gels \cite{ahmadi2006hydrodynamics,redford2024motor,Prost:2015la}, to cell assemblies or animal collectives \cite{duclos2018spontaneous,Blanch-Mercader:2018aa,Hakim:2017aa,Alert:2020aa}, as well as artificial active systems such as active colloids or robot swarms \cite{buttinoni2013dynamical,Bechinger:2016aa,Ben-Zion:aa}. 

A basic vital function of any living entity is to respond to environmental cues, typically to optimise the uptake of resources. 
Indeed, at the cellular level, motile cells have been shown to respond to various physicochemical cues, such as soluble or adherent chemical species, temperature, substrate stiffness, local structural order or topography \cite{yamada2019mechanisms}. 
In turn, on general grounds, a motile living entity is a source of local perturbations in its environment, because of either the release of soluble or adherent molecules, or mechanical forces involved in the migration mechanism \cite{zhou2014living,genkin2017topological,Gelimson:2016aa,dAlessandro:2021wx,clark2022self,lucas}. 
In the case of non-soluble cues, perturbations can  typically be localised and long-lived ; those will be hereafter called footprints. 

It was shown recently on various examples of cell types that the coupled dynamics of motile cells and their footprint fields can have striking consequences on the dynamics, both at the single cell scale \cite{dAlessandro:2021wx,Barbier-Chebbah:2022aa,lucas} and at the scale of cell collectives \cite{Gelimson:2016aa,li2017mechanism,Jacques2023}, such as ordering, aging, anomalous diffusion and even arrested dynamics; of note, similar behaviours were also recently observed in artificial active systems \cite{Hokmabad:2022aa,chen2024}.
In this article, inspired by these earlier works and in particular \cite{Jacques2023}, we propose a general theoretical framework to describe at the hydrodynamic level this broad class of systems where an active species (typically a cell monolayer, see Fig.\ref{fig:schem}) dynamically interacts with a  non-advected footprint field (typically a polymer matrix, see Fig.\ref{fig:schem}), which is  deposited by the active species.

We construct the  general hydrodynamics in the case in which the active species has polar or nematic aligning interactions, and show that the dynamic coupling to a footprint field can have striking consequences, which are absent in classical active hydrodynamics: (i) a symmetry-dependent modification to the isotropic-ordered transition, which can be either second or first order, (ii) a modification to spontaneous flow transitions, which can lead to oscillating flows and rotating fields, (iii) initial condition-dependent aging dynamics with long-lived transient states.

\textit{General hydrodynamic equations.} We consider a 2--dimensional, constant density, active fluid characterised by an orientational order parameter $A$, which typically describes the cell monolayer in the example of Fig.\ref{fig:schem}. We assume here, to remain general,  that $A$ is either the nematic order parameter: $A\equiv\Q=S(\hat{\bn}\otimes\hat{\bn}-\bsf{I}/2)$, where $\hat{\bn}=(\cos\theta,\sin\theta)$ is a unit vector, or the polar order parameter: $A\equiv\bp = p(\cos\theta,\sin\theta)$.
Similarly, the deposited footprint field $M$ may have nematic order, $M\equiv\M = S_M(\hat{\bfm}\otimes\hat{\bfm}-\bsf{I}/2)$ (where $\hat{\bfm} = (\cos\phi,\sin\phi)$), or polar order, $M\equiv\bfm = m(\cos\phi,\sin\phi)$.
Note that one can  also define nematic fields  from the polar order parameters: $\tilde{\bsf{P}} = \bp\otimes\bp-|\bp|^2\bsf{I}/2$ and $\tilde{\M} = \bfm\otimes\bfm - |\bfm|^2\bsf{I}/2$.

\begin{figure}
    \centering
    \includegraphics[width=0.48\textwidth]{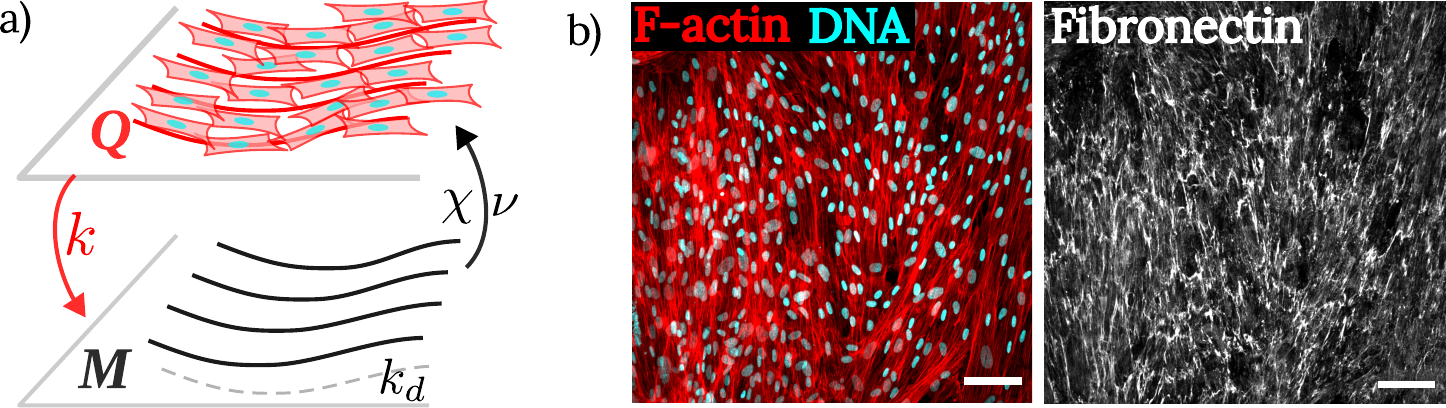}
    \caption{a) Cells and matrix interact through an active alignment of cells to the matrix (with coupling parameters $\chi$ and $\nu$), and deposition of matrix in the orientation of cells, with rate $k$. The matrix is also subject to uniform degradation with rate $k_d$. b) Fibroblasts plated on PAA gels, showing coherent alignment between the F-actin stress fibers of the cells (red, left panel), and the deposited fibronectin patterns (white, right panel). Scale bar 100$\mu$m. See \cite{Jacques2023} for experimental details.}
    \label{fig:schem}
\end{figure}

Following standard theories of active hydrodynamics \cite{Simha2002,Kruse2004,Kruse2005,marchetti2013hydrodynamics}, the dynamical equations for the active field $A$ is
\begin{equation}\label{dyn}
    D_t A = \mathcal{H}(A)/\gamma-\lambda \mathcal{E}(A) + \chi X(A,M),
\end{equation}
where $D_t$ is the corotational convective derivative (see supplementary material SM), $\mathcal{H}(A) = -\delta F/\delta A$ is the molecular field and $\gamma$ is a damping coefficient, $\lambda \mathcal{E}(A)$ is the flow-alignment coupling to the strain rate tensor $\E = (\nabla \bv + (\nabla \bv)^T-(\nabla\cdot\bv)\bsf{I})/2$ with parameter $\lambda$ \footnote{Note that for $A=\bp$, an extra coupling to $\bv$ --- and not its gradients --- exists for systems on substrates \cite{Brotto, Soni, Maitra2020}}, and $\chi X(A,M)$ represents the active coupling between $A$ and $M$ (to lowest order in a gradient expansion), see Table~\ref{tab}.
\begin{table}[]
    \centering
    \begin{tabular}{c|c|c|c|c}
         &  $\Q$--$\M$ & $\Q$--$\bfm$ & $\bp$--$\M$ & $\bp$--$\bfm$ \\
         \hline
       $X(A,M)$  & $\M$ & $\tilde{\M}$ & $\M\cdot\bp$ & $\bfm$\\
       $\mathcal{K}(A,M)$ & $k\Q$ & $k\Q\cdot \bfm$ & $k\tilde{\bsf{P}}$ & $k\bp$ \\
       $\mathcal{E}(A)$ & $\E$ & $\E$ & $\E\cdot\bp$ & $\E\cdot\bp$
       \end{tabular}
    \caption{Symmetry-dependent terms defined in Eqs.~\eqref{dyn}--\eqref{kinetic} for different combinations of fields: active nematic $\Q$, active polar $\bp$, nematic footprint $\M$, and polar footprint $\bfm$.}
    \label{tab}
\end{table}
We take a standard liquid crystal free energy and add to it aligning interactions with the footprint field: $F = \int_\Omega d\mathbf{r}\left[\frac{K}{2}(\nabla A)^2+\frac{\alpha}{2} A^2+\frac{\beta}{4} A^4 + f(A,M)\right]$, where $\partial f(A,M)/\partial A=-\nu X(A,M)$.
With this sign convention, $\chi,\nu>0$ represents an aligning interaction.

We write the momentum exchange between the active field and the substrate as
\begin{equation}\label{forcebalance}
    \xi\bv = \eta\nabla^2\bv - \nabla P + \nabla\cdot\sigma(A,\mathcal{H}),
\end{equation}
where $\bv$ is the velocity of the active field, $\eta$ its viscosity, $P$ is the  pressure that enforces the incompressibility of the flow ($\nabla\cdot \mathbf{v}=0$), and $\sigma(A,\mathcal{H})$ is the stress tensor  that contains contributions from $A$ \footnote{Note that for $A=\bp$,  an extra active force   $\propto \bp$ and an extra passive force $\propto \delta\mathcal{H}/\delta{\bf p}$ exists}.
Note that $\sigma(A,\mathcal{H})$ includes both the active stress and the  contributions of the corresponding passive liquid crystal theory (see SM); in particular  we distinguish the coupling $\chi$, of active origin, which does not appear in Eq.~\eqref{forcebalance} and $\nu$, which derives from a free energy and thus contributes to $\sigma(A,\mathcal{H})$ via $\mathcal{H}$. 

Finally, we take the dynamics of the footprint field $M$, assumed to be rigidly anchored to the substrate and thus non-advected, to obey the following first order kinetic equation:
\begin{equation}\label{kinetic}
\partial_t M = \mathcal{K}(A,M) - k_d M,
\end{equation}
where $\mathcal{K}(A,M)$ is the deposition rate (see Table~\ref{tab}), and $k_d$ is the intrinsic rate of degradation of the footprint, which is uniform and independent of orientation. Note that such deposition/ degradation dynamics need not be considered as active: it could be obtained by adding an appropriate term $\propto M^2$ into the free energy. Such equilibrium description, however, holds only for aligning interactions ($\nu>0$); as discussed below, the combination of non-aligning interactions $\nu<0$ with a positive deposition rate $k$ is nonreciprocal and thus inherently active. Finally, Eqs.~\eqref{dyn}--\eqref{kinetic} provide a minimal description of the dynamics of an active field $A$ coupled to a footprint field $M$ and form the core of this letter. They generalize \cite{Jacques2023} and share similarities with \cite{Adar2024}; we show below, using explicit examples, that the coupling to a footprint field can have striking consequences on the dynamics of the active fields. 

\if 
The model presented here is related to the one in \cite{Jacques2023}, although here we include both passive and active aligning interactions.
That article, motivated by experiments on cell layers, discussed the dynamics of topological defects, rather than the effect of a footprint field on alignment properties of active systems.
Compared to the framework introduced in Adar and Joanny, here we consider the constant cell density case, and do not neglect the passive liquid crystalline stresses present in the active field in the force balance equation. \AM{[AM: At this stage, it sounds cryptic because you haven't said why it is going to be important]} \AM{While \cite{Adar2024} neglected passive liquid crystalline stresses, we demonstrate that these modify the leading order in wavenumber physics in the presence of a footprint field.}
Compared to both works, we extend our treatment beyond the active nematic--nematic footprint to consider other combinations of symmetries.
\sbell{Ram's paper is actually 2d too.}
\AM{[AM: Why do we do that? We should say something regarding this.]}
\fi

\textit{Stationary, uniform phases : transition to order.} We first analyse the stationary, spatially uniform phases of the above theory. In the absence of a coupling to the footprint ($\chi=\nu=0$), the mean-field isotropic-ordered transition is classically continuous and occurs at $\alpha=0$ \footnote{Recall that nematic order is only quasi-long-ranged in two-dimensional systems}. The coupling to the footprint field  affects this transition in different ways, according to the symmetries of the active and footprint fields.
When the active field and its footprint share the same symmetry, the lowest order terms introduced by the couplings ($\chi,\nu$)  modify the isotropic-ordered transition threshold, but the  transition remains continuous. Indeed, 
Eq.~\eqref{kinetic} shows that at steady state, $kA=k_dM$, so the coupling terms $\propto M$ in the active field dynamics (Eq.~\ref{dyn}) renormalize the coefficient $\alpha$, shifting the transition to $\alpha-\gamma\tilde{\chi}k/k_d=0$, where $\tilde{\chi}=\chi+\nu/\gamma$ (see Fig.\ref{fig:order} and SM). This shows that the coupling to the footprint field can effectively mediate an enhanced aligning interaction of the active field and induce order, even for systems that would otherwise remain deep in the isotropic phase ($\alpha\gg 0$); in the context of cell monolayers, this suggests that nematic order can be induced by footprint interaction, even for weakly interacting cells (eg. typically low density), in agreement with \cite{li2017mechanism,Adar2024}.

For a mixed symmetry system, the footprint can even change the order of the transition. For example, in a $\bp$--$\M$ theory, Eq.~\eqref{kinetic} shows that $\M\propto \tilde{\bsf{P}}$, and so it is the quartic coefficient $\beta$ that is renormalized: the steady state magnitude of $\bp$ satisfies $-\alpha p+(\gamma\tilde{\chi}k/k_d-\beta)p^3-\kappa p^5=0$, where the $p^5$ term is included for stability. Here, $\gamma\tilde{\chi}k/k_d = \beta$ and $\alpha=0$ is a tricritical point with a second order transition when $\alpha < 0$ and $\gamma\tilde{\chi}k/k_d<\beta$ and a first order transition when $\alpha < 3(\gamma\tilde{\chi}k/k_d-\beta)^2/16\kappa$ and $\gamma\tilde{\chi}k/k_d >\beta$.
Strikingly, this suggests that the footprint interaction can induce a regime of coexistence of both isotropic and ordered phases, with a limit of metastability  given by $\alpha < (\gamma\tilde{\chi}k/k_d-\beta)^2/4\kappa$ (see Fig.~\ref{fig:order}).
\begin{figure}
    \centering
    \includegraphics[width=0.47\textwidth]{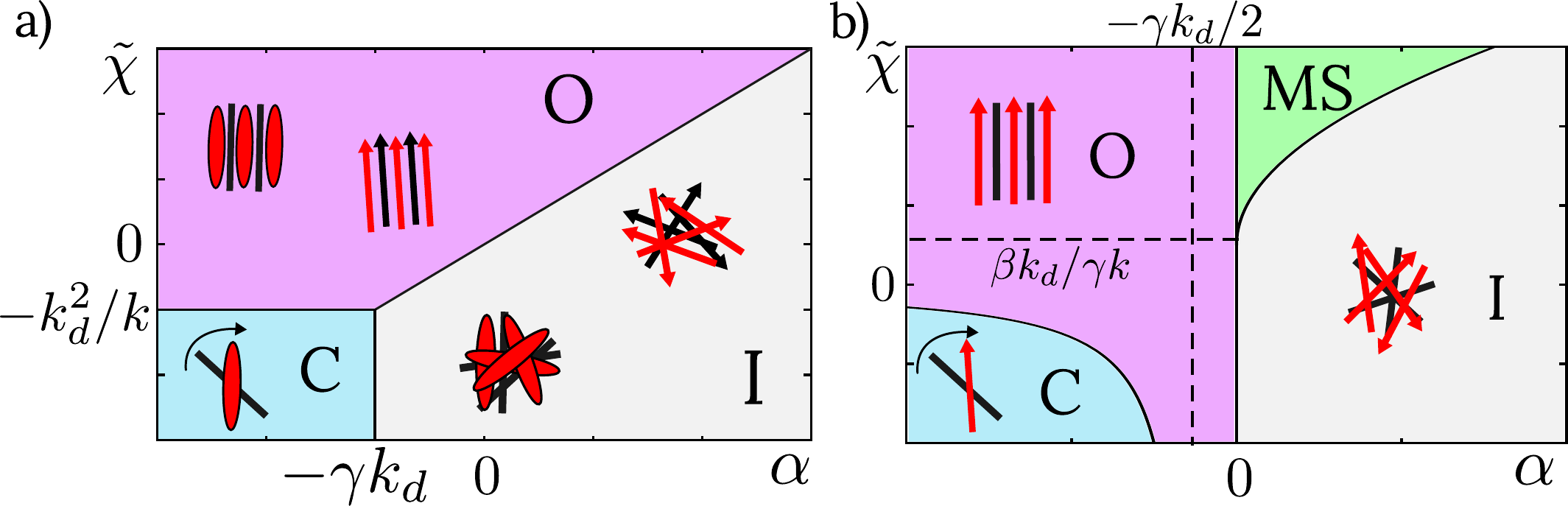}
    \caption{a)--b) Phase diagrams of uniform phases in the $\tilde{\chi}$--$\alpha$ plane for a) systems with a footprint of the same symmetry as the active field, and b) for a mixed symmetry $\bp$--$\M$ theory. The different phases are either ordered (O), isotropic (I), a coexistence regime (MS), or a rotating clock state (C). For the $\bp$--$\M$ theory, the clock state does not exist for $\alpha > -\gamma k_d/2$, and the tricritical point is situated at ($\alpha=0$, $\tilde{\chi}=\beta k_d/\gamma k$).}
    \label{fig:order}
\end{figure}

\textit{Non-uniform phases: flow instability.}
In the absence of a coupling to footprint ($\chi=\nu=0$) uniform static phases of active polar or nematic fluids have been shown to display linear instabilities, leading to spontaneously flowing states \cite{Simha2002,Voituriez2005,Giomi:2008aa,Marenduzzo:2007,Edwards2009,Pratley:2024aa,lavi2024} that play important roles in both artificial  and cellular systems~\cite{sanchez2012spontaneous,duclos2018spontaneous}.
We now analyse the impact of the coupling to a footprint field $\M$ on such spontaneous flow transitions for  an active nematic field  $\Q$. Eqs.~\eqref{dyn}--\eqref{kinetic} for this case explicitly read
\begin{gather}
    D_t \Q=\Hij/\gamma-\lambda\E + \chi\M \label{Qeq}\,,\\
    \partial_t \M = k\Q-k_d\M \label{Meq}\,,\\
    (\xi-\eta\nabla^2)\bv = -\nabla P - \zeta\nabla\cdot \Q + \bff_p\,,\label{Veq}
\end{gather}
where $\bff_p=(\nabla\Q):\Hij+\lambda\nabla\cdot\Hij + \nabla\cdot(\Q\cdot\Hij-\Hij\cdot\Q)$ is the force density due to passive liquid-crystalline stresses (see SM).
For simplicity, we  consider only the classical active stress $-\zeta \Q$, where $\zeta$ is a phenomenological coefficient\footnote{Upon solving for $\M$ in terms of $\Q$, the active alignment term $\chi\M$ in \eqref{Qeq} becomes $\chi(k/k_d)\Q$. Active contributions of this form were discussed in~\cite{Kruse2005,julicher2007active}}. 

We consider the linear stability of a uniform non-flowing aligned state in infinite space, with $\theta = \phi = 0$, and $S_M/S=k/k_d$. For simplicity, we assume that the system remains deep in the nematic phase, so that $S\equiv 1$, even upon perturbation (see SM). 
In Fourier space, where $\tilde{f}(\mathbf{q},t) = \int d\mathbf{r}\; f(\mathbf{r},t) e^{-i\mathbf{q}\cdot\mathbf{r}}$,  the linearised joint dynamics of $\tilde{\Xi}=\tilde{\theta}+\tilde{\phi}$ and the relative angular separation $\tilde{\Delta} = \tilde{\theta} - \tilde{\phi}$ reads
\begin{gather}
  \partial_t\tilde{\Xi} =-\frac{\tilde{K}_\bq+\tilde{\zeta}_\bq}{2}\tilde{\Xi} -\left( \frac{\tilde{\chi}_\bq k}{k_d}-k_d+\frac{\tilde{K}_\bq +\tilde{\zeta}_\bq}{2}\right)\tilde{\Delta} \label{eq:stabTh}\,,\\
  \partial_t\tilde{\Delta} = -\left(\frac{\tilde{\chi}_\bq k}{k_d}+k_d+\frac{\tilde{K}_\bq+\tilde{\zeta}_\bq}{2}\right)\tilde{\Delta}-\frac{\tilde{K}_\bq+\tilde{\zeta}_\bq}{2}\tilde{\Xi} \,.\label{eq:stabDelt}
\end{gather}
Representing the wavevector in polar coordinates, $\bq=q(\cos\psi,\sin\psi)$, $\tilde{K}_\bq = K q^2\left[\frac{1}{\gamma}+\frac{q^2(\lambda\cos 2\psi-1)^2}{2(\xi+\eta q^2)}\right]$, is a positive quantity that suppresses instability, stemming from the Frank elasticity of the active nematic; $\tilde{\zeta}_\bq=\frac{q^2(\lambda\cos 2\psi-1)\cos 2\psi}{2(\xi+\eta q^2)}(\zeta+\lambda\gamma\chi(k/k_d))$ can be interpreted  as the contribution of the active stress that can trigger  instability, and  finally $\tilde{\chi}_\bq=\chi + \left[\frac{1}{\gamma}+\frac{q^2(\lambda\cos 2\psi-1)^2}{2(\xi+\eta q^2)}\right]\nu$ is specific to the footprint field as it vanishes for $\chi=\nu=0$. Of note, besides $\tilde{\chi}_\bq$, footprint couplings 
thus enter the dynamics only by renormalizing the active stress parameter $\zeta\to \zeta + \lambda\gamma\chi(k/k_d)$ in $\tilde{\zeta}_\bq$. This illustrates the intrinsic active nature of the $\chi$ coupling term  discussed above, which cannot be derived consistently from a free energy, as opposed to the passive coupling $\nu$. Importantly, the linear system of Eqs. \eqref{eq:stabTh},~\eqref{eq:stabDelt} fully determines the stability of the static nematic phase ; writing $\tilde{\Xi},\tilde{\Delta}\propto e^{s_\bq t}$ it is dictated by the sign of its  eigenvalues $s_\bq$, which can be obtained analytically (see SM). We discuss below  the impact of the footprint on stability, and focus on the large scale ($q\to 0$) regime.  

We  focus on the general case of non-vanishing friction ($\xi=0$ is discussed in SM). Because, in this case, $\tilde{\zeta}_\bq,\tilde{K}_\bq =\mathcal{O}(q^2)$, Eqs. \eqref{eq:stabTh}--\eqref{eq:stabDelt} yield a separation of timescales with $\tilde{\Xi}$ being a slow variable (with $\mathcal{O}(q^2)$ relaxation rate), whereas  $\tilde{\Delta}$ is fast (with $\mathcal{O}(1)$ relaxation rate). The dynamics of the slow variable can then be written as
\begin{equation}
    \partial_t\tilde{\Xi}=-(\tilde{K}^0_\bq+\tilde{\zeta}^0_\bq)\frac{k_d}{k_d+\tilde{\chi}_\bq^0 k/k_d}\tilde{\Xi}\,,\label{eq:stabTh2}
\end{equation}
 where $\tilde{\zeta}^0_\bq,\tilde{K}^0_\bq, \tilde{\chi}^0_\bq$ denote the leading order of the corresponding quantities for $q\to 0$. Because $\tilde{\chi}^0_\bq,k_d,k>0$, this shows that the uniform state is unstable (with $s_\bq\sim q^2$) for $\tilde{K}^0_\bq+\tilde{\zeta}^0_\bq<0$,  which yields the condition:
\begin{equation}\label{eq:stabcond}
    \frac{2K\xi}{\gamma} +\cos2\psi(\lambda\cos 2\psi-1)(\zeta+\lambda\gamma\chi(k/k_d))<0.
\end{equation}
Taking $\chi=0$, one recovers the case without footprint studied in \cite{Maitra2018}. The footprint thus modifies the \textit{threshold} of the instability only through the renormalization of the active stress by the active alignment  $\zeta\to\zeta + \lambda\gamma\chi(k/k_d)$.
The passive alignment parameter, $\nu$, and the footprint deposition and degradation rates $k,k_d$, modify the magnitude of the instability's \textit{growth rate} $s_{\bq}$, and slow the dynamics down, as quantified by Eq.~\eqref{eq:stabTh2}.
For  flow-aligning fluids---$|\lambda|>1$, reported to be the case of epithelial cells \cite{aigouy2010,blanch2021}---because $\lambda\cos 2\psi(\lambda \cos 2\psi - 1) > 0$ for all wavevector directions $\psi$,  the footprint-induced active stress stabilizes the system to both splay ($\lambda>0)$ and bend ($\lambda<0$) instabilities.
Conversely, for flow-tumbling systems, $|\lambda| < 1$, it is always possible to find a value of $\psi$ such that $\lambda\cos 2\psi(\lambda\cos 2\psi - 1)< 0$, i.e., the active footprint coupling acts to promote instability.

Such spontaneous flow transition can be conveniently illustrated in confined semi-infinite slab geometry, such as in~\cite{Voituriez2005,Edwards2009,Bell2022}, where the system is assumed to be invariant along the infinite direction.
We take the initial state aligned parallel ($\theta_0=\pi/2$) to the infinite $y$-direction, which sets $\psi = \pi/2$ and allows for a splay instability, see Fig.~\ref{fig:flow}a) (the bend instability ($\theta_0=0$, $\psi = 0$), can be obtained with the transformations $\zeta\to -\zeta$ and $\lambda \to -\lambda$). In this geometry, the analysis of $s_\bq$ up to $\mathcal{O}(q^4)$ defined by Eqs.~\eqref{eq:stabTh}--\eqref{eq:stabDelt} (see SM) yields the critical slab width $L_c$ at which the active nematic starts to flow:
\begin{equation}\label{eq:stabStrip}
    \left(\frac{\pi}{L_c}\right)^2=-\frac{(\zeta+\lambda\gamma\chi(k/k_d))\gamma(\lambda+1)}{2K\tilde{\eta}}-\frac{\xi}{\tilde{\eta}}\,,
\end{equation}
where $\tilde{\eta}=\eta+\gamma(\lambda+1)^2/2$.
This confirms that the active coupling to footprint $\chi$ stabilizes the uniform state and delays the onset of flows for flow-aligning systems $|\lambda|>1$. Strikingly, for flow-tumbling systems, the uniform state can be unstable for pure splay perturbations for $-1<\lambda<0$, even in the absence of active stress $\zeta = 0$ (see Fig.~\ref{fig:flow}b). For a pure bend perturbation, $\theta_0=0$, this destabilisation occurs for $0<\lambda<1$. Finally, this shows that the active coupling to footprint $\chi$ induces an effective active stress $-\lambda\gamma\chi(k/k_d) \Q$ that  shifts the classical active coupling $\zeta$ to $\zeta+\lambda\gamma\chi(k/k_d)$ (thus in a $\lambda$-dependent manner), thereby directly impacting the onset of spontaneous flows. 

\begin{figure}
	\centering
	\includegraphics[width=0.45\textwidth]{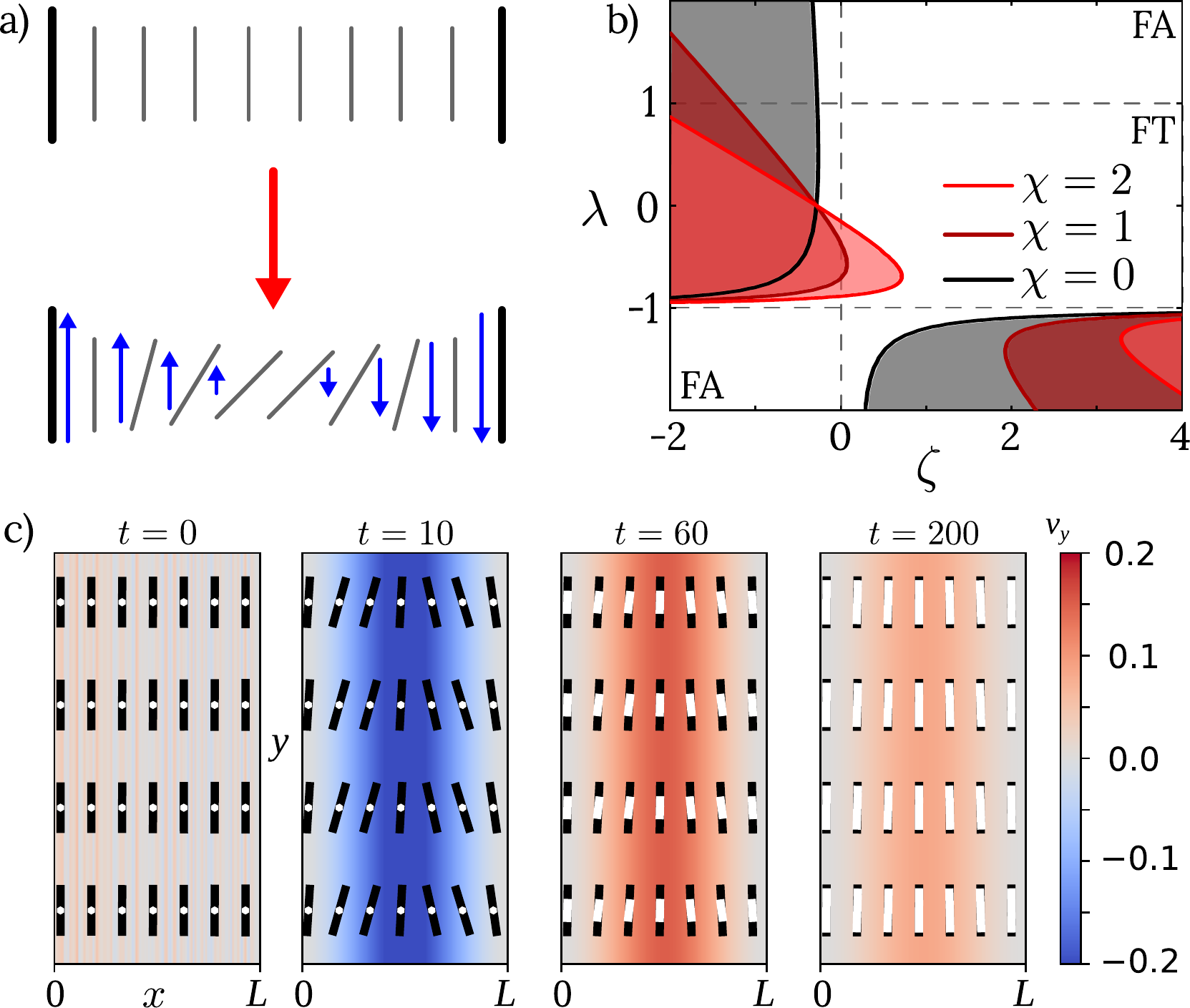}
	\caption{a) Schematic of the activity-induced splay instability for a slab geometry with $\theta_0=\pi/2$ and free-slip strong anchoring boundary conditions. The nematic orientation is indicated by the gray lines, and the resulting shear flow by the blue arrows. b) The stability in $\zeta$--$\lambda$ space of the slab geometry displayed in a), as calculated analytically from the sign of the growth rates $s_\bq$. The shaded regions indicate instability for different values of the active alignment $\chi$. The slab width $L$ is fixed. c) Transient flows, order parameter (black bars) and footprint field (white bars) for an initially uniform state without footprint ($\theta(x,0) = \pi/2$, $\M(t=0)=\mathbf{0}$) under no-slip strong anchoring boundary conditions. The contractility $\zeta$ is chosen so that the state without footprint flows spontaneously, but that the effective activity with footprint, $\zeta+\lambda\gamma\chi(k/k_d)$, has a stable uniform steady state (according to Eq.~\eqref{eq:stabStrip}). See also Supp. Movie 1. Numerical resolution scheme and values of parameters for all panels are given in SM.} 
	\label{fig:flow}
\end{figure}
 Of note, the above analysis excludes the singular limit $k=k_d=0$, which can be obtained as a byproduct from Eqs.~\eqref{eq:stabTh}--\eqref{eq:stabDelt}. This  corresponds to a constant aligning field acting on the active nematics, which can have broad applications. In the context of cell monolayers, this corresponds to a substrate with fixed structural nematic order, which can occur \emph{in vivo} because of the extra-cellular matrix organisation, or can be created \emph{in vitro} by micropatterning techniques~\cite{Leclech2022}; more generally aligning fields could be realized by chemical gradients~\cite{Ibrahimi2023}, strain~\cite{De2007}, strain-rate \cite{Muhuri}, electric, or curvature fields \cite{Bell2022}. The analysis (see SM) shows that for $\xi\not=0$ the fixed footprint stabilizes the long wavelength dynamics ($s_{\bq=0}<0$), in contrast to the above case of dynamic footprint, and thus restricts the instability to a finite range of wavelengths.

\textit{Transient flows and aging.}
For times $t \ll 1/k_d$, the coupling to the footprint field induces effective memory effects in the dynamics of the  active field. Before eventually reaching the different possible steady states described above, the system  displays transient, aging dynamics that depend on initial conditions and can be long-lived in the limit $k_d\to 0$. These transient behaviors are strikingly different from classical relaxation dynamics to either the isotropic, ordered, or flowing states. To  illustrate this effect, we consider the above slab geometry with uniform nematic order ($\theta_0=\pi/2$), in absence of footprint ($M=0$) at $t=0$; this would be typically the case of a cell monolayer invading a fresh environment, or when footprint deposition is triggered by an external cue. There exist regions of parameter space where the bare contractility $\zeta$ induces a flow transition, whereas the renormalized contractility $\zeta + \lambda\gamma\chi(k/k_d)$ leads to a stable uniform steady state (see Fig.\ref{fig:flow}), and can even be of opposite sign.
In this case, for times $t\ll 1/k$, the system  undergoes a classical splay instability and starts to flow. Eventually, at times $t\gtrsim 1/k_d $, the system undergoes a flow reversal before relaxing to the uniform state with vanishing flows, see Fig.~\ref{fig:flow}c and movie 1). This shows that the coupling to footprint can at long timescales induce  an effective freezing of the dynamics, which could provide an alternative mechanism to the observed dynamical arrest in cell monolayers, often attributed to active jamming~\cite{Bi2016,Garcia2015,Hakim:2017aa}.

\textit{Anti-aligning footprints: non-reciprocal interactions.}
We now extend the discussion to systems where the footprint is deposited in alignment with the active field, but the active field naturally antialigns with its footprint either via the active or passive coupling, so that $\tilde \chi<0$. This leads to non-reciprocal interactions between the active and footprint fields \cite{Aranson_LLC_PNAS,fruchart2021non}. Note that, because it requires that $\tilde \chi$ and $k$ to have opposite signs, this could not be realised in equilibrium systems, but can arise in generic active materials.
\begin{figure}
\centering
\includegraphics[width=0.48\textwidth]{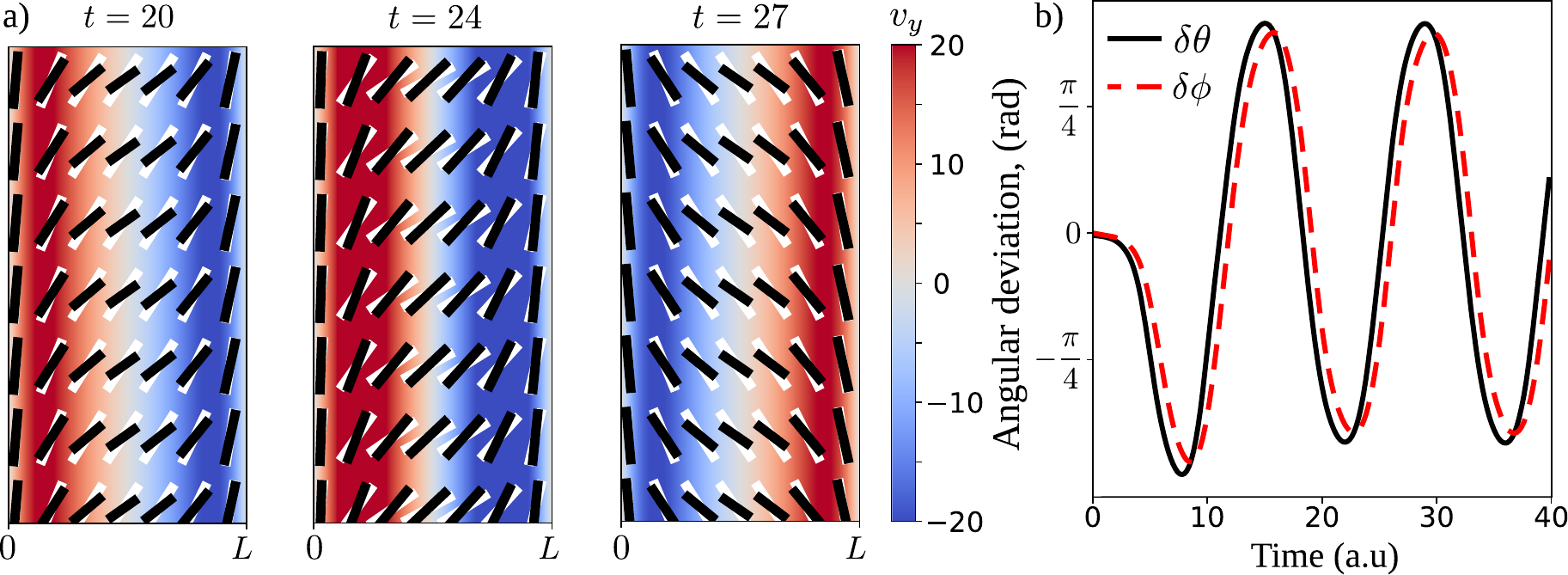}
\caption{Numerical snapshots of a quasi-1D oscillating shear state of an active nematic (black bars) with a nematic footprint (white bars) in a thin strip with no-slip strong anchoring boundary conditions of $\theta=\pi/2$ at time $t=\tau$ (a) and a half-period later, $t=\tau + T/2$ (b). c) The deviation from $\pi/2$ of the two director fields in the center of the strip as a function of time: the active nematic ($\delta\theta=\theta(L/2)-\pi/2$) and the footprint ($\delta\phi = \phi(L/2)-\pi/2$). See also Supp. Movie 2.}
\label{fig:hopf}
\end{figure}
Eqs. \eqref{Qeq}, \eqref{Meq}, \eqref{Veq} show that the velocity field affects the order parameter dynamics at $\mathcal{O}(q^2)$ in  systems with non-vanishing friction, and is thus sub-dominant at large scales ($q\to 0$). Therefore, we first neglect flow interactions and  seek non-flowing, uniform steady states. The analysis of \eqref{Qeq},\eqref{Meq} (see SM) shows that a $\Q$--$\M$ system
can form a steadily rotating state with $\cos 2\Delta_0=k_d\sqrt{-1/k\tilde{\chi}}$, $\Delta_0\neq 0, \pi/2,\pi$, where $\Delta=\theta-\phi$ is the angular separation between the directors of the two order parameters, for parameters values given in Fig.~\ref{fig:order}a \footnote{A similar  steadily-rotating nematic state due to antagonistic alignment interactions between two nematic fields    \cite{zhou2014living} was first discussed by M. E. Cates, S. Ramaswamy and E. Tjhung (unpublished)}.
In this state, the two director fields rotate in the same (arbitrary) direction (resulting from spontaneous symmetry breaking), with the footprint director $\phi$ lagging behind the active director $\theta$: $\theta=\omega t$, $\phi=\omega t-\Delta_0$, where $\omega=\pm\sqrt{-(\tilde{\chi}k+k_d^2)}/2$.
Even though it emerges from a different mechanism---repulsive coupling to footprint---such rotating state is analogous to the time cholesteric state discussed in \cite{AM_chi, Callan-Jones}. 

Remarkably, in the semi-infinite slab geometry discussed above, with the boundary conditions $\theta(0)=\theta(L)=\pi/2$, geometric confinement forbids steadily rotating states and a numerical resolution of \eqref{Qeq}, \eqref{Meq}, \eqref{Veq} shows the existence of a non-uniform oscillating state with a corresponding oscillatory shear flow as shown in Fig.~\ref{fig:hopf} and movie 2; it is shown in the SM that such oscillatory dynamics can be interpreted as a Hopf bifurcation of the ($\theta$, $\phi$) dynamics. Note that while comparable to \cite{Mori2023}, oscillating flows have here a different origin. This further illustrates that the coupling to footprint can induce a broad variety of dynamical patterns.

\textit{Conclusion}
Inspired by recent experimental observations, we propose a general hydrodynamic theory that describes the  dynamics of a polar or nematic active field interacting with a footprint field it deposits. This dynamic coupling to a footprint field induces remarkable effects absent in classical active hydrodynamics, which could have important implications for cellular systems. First, the footprint deeply changes  the isotropic-ordered transition by favoring the ordered phase suggesting that nematic cell monolayers could be observed even for weakly interacting (typically low density) systems. Second, footprints modify the threshold to flow instabilities and their dynamics, and can even induce new dynamical phases, such as rotating or oscillating states.  
Last, the dynamic footprint deposition induces effective memory effects, with long lived transients that depend on initial conditions, and thus effectively induces aging behaviours in cellular systems; in particular for flow-aligning systems (which was found to be the case of cell monolayers) a slow footprint deposition can suppress initially flowing states and thus effectively freeze the dynamics, which could provide an alternative mechanism to the dynamical arrest  of cell monolayers observed in different contexts~\cite{Bi2016,Garcia2015}. Finally, our results show that footprint deposition could be an important mechanism that plays a key role in determining a broad range of  dynamical phases of cellular systems, or more generally active systems inducing long-lived perturbations to their environment.


\textit{Acknowledgements} We thank C. Jacques and D. M. Vignjevic for the experimental panels in Fig.~\ref{fig:schem}b), and R. Adar and JF. Joanny for discussions and sharing  \cite{Adar2024}. Support from ERC synergy grant Shapincellfate is acknowledged.

\newpage
\clearpage

\title{Supplementary Material: Ordering, spontaneous flows and aging in active fluids  depositing tracks}
\maketitle
\onecolumngrid
\appendix
\setcounter{equation}{0}

\section{Definition of symmetry-dependent terms}
\subsection{Corotational convective derivative}
The corotational convective derivative is defined differently for active nematic or polar fields:
\begin{gather}
	\frac{D \Q}{Dt} \equiv \partial_t \Q + (\bv \cdot \nabla)\Q + \Om\cdot \Q - \Q\cdot \Om \, ,\\
	\frac{D \bp}{Dt} \equiv \partial_t \bp + (\bv \cdot \nabla)\bp + \Om \bp \, ,
\end{gather}
where $\bv$ is the flow velocity, and $\Omega_{ij}=(1/2)(\partial_iv_j-\partial_jv_i)$ is the vorticity tensor.
\subsection{Stress tensor}
Following standard theories of active gels (for a review see \cite{marchetti2013hydrodynamics}), the stress tensor for an active nematic is written:
\begin{equation}
	\sigma \equiv -\zeta\Q  + \lambda \Hij + \Q\cdot\Hij - \Hij\cdot\Q + \sigma^E\, ,
\end{equation}
where $\Hij = -\delta F/\delta\Q$ is the molecular field ($F$ is a free energy). The Ericksen stress, $\sigma^E$, is written in index notation:
\begin{equation}
	\sigma^E_{ij} = -\frac{\partial f}{\partial \nabla_i Q_{kl}}\nabla_j Q_{kl}\, ,
\end{equation}
where $f$ is the free energy density such that $F = \int f\; d^2\br$. 

For an active polar gel, the stress tensor is written:
\begin{equation}
	\sigma \equiv - \zeta(\bp \otimes \bp - \nabla |\bp|^2 \bsf{I}/2) + \frac{\lambda}{2}\left(\bp\otimes\bh + \bh \otimes\bp\right)-\frac{1}{2}\left(\bp \otimes \bh - \bh \otimes \bp\right) + \sigma^E\, ,
\end{equation}
where $\bh = -\delta F/\delta \bp$ is the molecular field, and where the Ericksen stress is written:
\begin{equation}
	\sigma^E_{ij} = -\frac{\partial f}{\partial \nabla_i p_k}\nabla_j p_k\, .
\end{equation}
\section{Transition to order}
\subsection{Active nematic fluid with a nematic footprint}
For an active nematic $\Q$ coupling to its nematic footprint $\M$, neglecting elastic and flow-alignment contributions, Eqs.~(1)~\&~(2) are written:
\begin{gather}
	\partial_t \Q = -\frac{\alpha}{\gamma}\Q - \frac{2\beta}{\gamma}\text{Tr}[\Q^2]\,\Q +\tilde{\chi}\M \label{eq:dtQ}\,,\\
	\partial_t \M = k\Q-k_d \M\,, \label{eq:dtM}
\end{gather}
where $\tilde{\chi}=\chi + \nu/\gamma$.
At steady state, Eq.~\eqref{eq:dtM} imposes $k\Q=k_d \M$: $\M$ renormalizes $\alpha$ such that the active nematic order parameter $S$ satisfies $(\gamma\tilde{\chi}k/k_d-\alpha)S-\beta S^3=0$.
This permits a non-zero $S$ when $\gamma\tilde{\chi}k/k_d>\alpha$.
The Jacobian for the isotropic state, $J(0,0)=\left(\begin{smallmatrix} -\alpha/\gamma & \tilde{\chi} \\ k & -k_d\end{smallmatrix}\right)$,
has a positive eigenvalue for $\tilde{\chi}\gamma k/k_d>\alpha$, so the isotropic--ordered transition remains continuous within the linear theory: the footprint interaction modifies the threshold.

Note that, in equilibrium, both the dynamics of $\Q$ and $\M$ should minimise a free energy. The positive-definiteness of Onsager dissipative coefficients then implies that $\tilde{\chi}$ and $k$ must have the same sign; no such restriction applies in an active system.

\subsection{Active polar fluid with a nematic footprint}
Once again neglecting elastic and flow-alignment contributions, the dynamics for a polar active field $\bp$ coupled to its nematic footprint $\M$ is written: 
\begin{gather}
	\partial_t \bp = -\frac{\alpha}{\gamma}\bp - \frac{\beta}{\gamma}\bp^3 + 2\tilde{\chi}\M\cdot\bp - \frac{\kappa}{\gamma} \bp^5\,, \\
	\partial_t \M = k\tilde{\bsf{P}}-k_d\M\,,
\end{gather}
recalling that $\tilde{\bsf{P}} = \bp\otimes\bp-|\bp|^2\bsf{I}/2$.
Since $\M\sim \bp^2$, the footprint does not affect the linear stability of the isotropic state.
Instead, notice that the dynamics for the magnitude of $\bp$ can be written as the derivative of a potential, $\partial_t p = -(1/\gamma)d V(p)/dp$, where the effective potential $V(p)$ is written
\begin{equation}
	V(p) = \frac{\alpha p^2}{2}-\frac{(\gamma\tilde{\chi}k/k_d-\beta)p^4}{4}+\frac{\kappa p^6}{6}\,.
\end{equation}
As is standard, the ordered state can be found by solving for $dV(p)/dp=0$:
\begin{equation}
	p_0^2 = \frac{\gamma\tilde{\chi}k/k_d-\beta + \sqrt{(\gamma\tilde{\chi}k/k_d-\beta)^2-4\alpha\kappa}}{2\kappa}\,,
\end{equation}
from which it is clear that the ordered state with order parameter value $p_0\neq 0$ becomes metastable for $\gamma\tilde{\chi}k/k_d>\beta$ and $\alpha < (\gamma\tilde{\chi}k/k_d-\beta)^2/4\kappa$.
There is a tricritical point at $\alpha=0$ and $\gamma\tilde{\chi}k/k_d=\beta$.
Again, as is standard, the ordered state becomes \textit{globally} stable when $V(p_0)=0$. Calculating this, we find that the system undergoes a first order phase transition when $\alpha < 3(\gamma\tilde{\chi}k/k_d-\beta)^2/16\kappa$ and $\gamma\tilde{\chi}k/k_d>\beta$.

\section{Spontaneous flow transitions}
The $\Q$--$\M$ theory is written:
\begin{gather}
	\partial_t \Q + (\bv\cdot\nabla)\Q+\Om\cdot\Q-\Q\cdot\Om= \frac{\Hij}{\gamma}-\lambda\E + \chi \M  \label{eq:Qdyn}\,,\\
	\partial_t\M = k\Q-k_d\M \label{eq:Mdyn}\,,\\
	\xi\bv = -\nabla P - \zeta\nabla\cdot\Q + \eta\nabla^2\bv - \Hij\cdot\nabla\Q + \lambda\nabla\cdot\Hij + \nabla\cdot(\Q\cdot\Hij-\Hij\cdot\Q) \label{eq:fb}\,,
\end{gather}
where $\Hij\cdot\nabla\Q$ is $H_{kl}\nabla_iQ_{kl}$, for $\Q=S(\hat{\bn}\otimes\hat{\bn}-\bsf{I}/2)$, $\hat{\bn}=(\cos \theta,\sin\theta)$, $\E = (\nabla \bv + (\nabla \bv)^T-(\nabla\cdot\bv)\bsf{I})/2$, where $\nabla{\bf v}\equiv\partial_iv_j$ and $\Omega_{ij}=(1/2)(\partial_iv_j-\partial_jv_i)$. The molecular field $\Hij = -\delta F/\delta\Q$ is the functional derivative of a standard nematic free energy $F$:
\begin{equation}
	\int_S d\br \left[\frac{K}{2}(\nabla\Q)^2+\frac{\alpha}{2}\Tr\Q^2 + \frac{\beta}{2}(\Tr\Q^2)^2-\nu\Q:\M\right].
\end{equation}
We will consider the stability of the nematic phase, i.e., when $S_0=\sqrt{[-\alpha+(\nu+\gamma\chi)(k/k_d)]/\beta}$ is real.

\section{2D stability analysis}

We first rewrite the system in terms of the order parameter amplitudes ($S$, $S_M$), and the angular dynamics ($\theta$, $\phi$). The time evolution of the order parameters amplitudes are:
\begin{gather}
	\begin{split}
		\partial_t S = \frac{1}{\gamma}\left[K\nabla^2 S - (\alpha + \beta S^2)S\right]&-\frac{\lambda}{2}\left[(\partial_x v_x-\partial_y v_y)\cos 2\theta + (\partial_x v_y + \partial_y v_x)\sin 2\theta)\right] \\ &+\left(\chi+\frac{\nu}{\gamma}\right)S_M\cos 2(\theta-\phi), \end{split} \\
	\partial_t S_M = kS\cos 2 (\theta-\phi) - k_d S_M.
\end{gather}
We linearise these equations about an aligned steady state with $\theta_0=\phi_0 = 0$, $S_{M,0}=k/k_d S_0$, $S_0=\sqrt{|\bar{\alpha}|/\beta}$, with $\bar{\alpha}=\alpha-(\chi\gamma+\nu)k/k_d$:
\begin{gather}
	\partial_t \delta S = \frac{1}{\gamma}\left[K\nabla^2 \delta S - (\alpha + 3\beta S_0^2)\delta S\right]-\frac{\lambda}{2}\left(\partial_x v_x - \partial_y v_y\right)+\left(\chi+\frac{\nu}{\gamma}\right)\delta S_M, \\
	\partial_t \delta S_M = k\delta S-k_d \delta S_M.
\end{gather}
Importantly, we will see later that, in systems on substrates, the velocity field contributes terms at $\mathcal{O}(q^2)$. Therefore, to $\mathcal{O}(q^0)$, the dynamics of the order parameter amplitudes decouple from that of the angle and become
\begin{gather}
	\partial_t \delta S = -\frac{1}{\gamma}\left(\alpha + {3}|\bar{\alpha}| \right){\delta}S+\left(\chi+\frac{\nu}{\gamma}\right){\delta}S_M, \\
	\partial_t {\delta}S_M = k\delta S-k_d\delta S_M.
\end{gather}
Assuming $\alpha<0$, the first equation becomes
\begin{equation}
	\partial_t \delta S = -\frac{1}{\gamma}\left[2|\alpha| + {3}\left(\chi\gamma+\nu\right)\frac{k}{k_d} \right]{\delta}S+\left(\chi+\frac{\nu}{\gamma}\right){\delta}S_M.
\end{equation}
The coupled equations for $\delta S$ and $\delta S_M$ implies the eigenvalues
\begin{equation}
	\kappa^{SS_M}_\pm=-\left[\frac{k_d}{2}+\frac{2|\alpha|k_d+3k(\nu+\gamma\chi)}{k_d\gamma}\pm\sqrt{\left(\frac{k_d}{2}-\frac{2|\alpha|k_d+3k(\nu+\gamma\chi)}{k_d\gamma}\right)^2+\frac{4k}{\gamma}(\nu+\gamma\chi)}\right].
\end{equation}
These eigenvalues are stabilising when $k_d|\alpha|+k(\nu+\gamma\chi)>0$; this implies that small $q$ amplitude fluctuations always decay in the ordered state. Since the fluctuations of the order parameter amplitudes have a finite decay rate at vanishingly small wavenumbers, from here on, we will examine dynamics beyond timescales at which these fluctuations decay and replace $S$ and $S_M$ by their steady state values.


The footprint's angular dynamics is written:
\begin{equation}
	\partial_t \phi= \frac{k S}{2S_M}\sin 2(\theta-\phi)\, ,
\end{equation}
which, when linearized, gives:
\begin{equation}
	\partial_t \phi = k_d (\theta-\phi).
\end{equation}
The angular dynamics for the active field is written:
\begin{multline}
	\partial_t \theta = \frac{K}{\gamma}\nabla^2\theta + \frac{\lambda}{2S}\sin 2\theta(\partial_x v_x - \partial_y v_y) + \frac{1}{2}\left(1-\frac{\lambda}{S}\cos2\theta\right)\partial_x v_y +\frac{1}{2}\left(-1-\frac{\lambda}{S}\cos2\theta\right) \partial_y v_x \\
	-\left(\chi+\frac{\nu}{\gamma}\right)\frac{S_M}{2S}\sin 2(\theta-\phi), \label{SMeq:dtTh}
\end{multline}
which, when linearized, yields
\begin{equation}
	\partial_t \theta = \frac{K}{\gamma}\nabla^2\theta + \frac{1}{2}\left(1-\frac{\lambda}{S_0}\right)\partial_x v_y +\frac{1}{2}\left(-1-\frac{\lambda}{S_0}\right) \partial_y v_x 
	-\left(\chi+\frac{\nu}{\gamma}\right)\frac{k}{k_d} (\theta-\phi). \label{SMeq:dtTh2}
\end{equation}


Fixing $S$ and $S_M$ to their steady-state values, the force balance equation takes the form
\begin{equation}
	(\xi-\eta\nabla^2)\mathbf{v} = -\nabla P - S_0\zeta\begin{pmatrix} \partial_y \theta \\ \partial_x\theta\end{pmatrix}+\lambda\begin{pmatrix}\partial_y \\ \partial_x \end{pmatrix}(K S_0\nabla^2\theta - [\alpha + \beta S_0^2]S_0\theta + \nu S_{M,0}\phi)+S_0^2\begin{pmatrix}\partial_y h_\theta \\ -\partial_x h_\theta\end{pmatrix},
\end{equation}
where $h_\theta=K\nabla^2\theta - \nu(k/k_d)(\theta-\phi)$. Using the expression of $S_0$, this becomes 
\begin{equation}
	(\xi-\eta\nabla^2)\mathbf{v} = -\nabla P - S_0(\zeta+\lambda\gamma\chi(k/k_d))\begin{pmatrix} \partial_y \theta \\ \partial_x\theta\end{pmatrix}+S_0^2\begin{pmatrix}(\lambda/S_0+1)\partial_y h_\theta \\ (\lambda/S_0 -1)\partial_x h_\theta\end{pmatrix}.
\end{equation}
Projecting ${\bf v}$ transverse to the wavevector in Fourier space using the projection operator $\delta_{ij}-q_iq_j/q^2$, we obtain the velocity field
\begin{equation}
	\tilde{\mathbf{v}} = \frac{iS_0q}{\xi+\eta q^2}\left[-\{S_0(\zeta+\lambda\gamma\chi (k/k_d))\cos 2\psi\}\tilde{\theta}+S_0^2\left(\frac{\lambda}{S_0}\cos 2\psi - 1\right)\tilde{h}_\theta\right]\begin{pmatrix}-\sin\psi \\ \cos\psi \end{pmatrix},
\end{equation}
where $q_x=q\cos\psi$, and $q_y=q\sin\psi$.
Substituting this into the Fourier transform of Eq.~\eqref{SMeq:dtTh} gives
\begin{equation}
	\partial_t \tilde{\theta} = -[\tilde{K}_\bq+\tilde{\zeta}_\bq]\tilde{\theta}-\tilde{\chi}_\bq(k/k_d)(\tilde{\theta}-\tilde{\phi}),
\end{equation}
where
\begin{gather}
	\tilde{K}_\bq =Kq^2\left[\frac{1}{\gamma}+\frac{S_0^2q^2}{2(\xi+\eta q^2)}\left(\frac{\lambda}{S_0}\cos 2\psi-1\right)^2\right]\,, \\
	\tilde{\zeta}_\bq = \left(\frac{\lambda}{S_0}\cos 2\psi-1\right)\frac{S_0q^2}{2(\xi+\eta q^2)}(\zeta+\lambda\gamma\chi(k/k_d))\cos2\psi\,, \\
	\tilde{\chi}_\bq = \chi + \left[\frac{1}{\gamma}+\left(\frac{\lambda}{S_0}\cos 2\psi-1\right)^2\frac{S_0^2q^2}{2(\xi+\eta q^2)}\right]\nu\,.
\end{gather}

We assume that we are deep in the nematic state, such that $|\alpha|,\beta\gg 1$, and for simplicity, we set $\alpha = -\beta$, such that $S_0 = 1$ without loss of generality. Then, up to linear terms in the activity and footprint alignment parameters, we recover the quantities as defined in the main text:
\begin{gather}
	\tilde{K}_\bq =Kq^2\left[\frac{1}{\gamma}+\frac{q^2\left(\lambda\cos 2\psi-1\right)^2}{2(\xi+\eta q^2)}\right], \\
	\tilde{\zeta}_\bq = \frac{q^2\left(\lambda\cos 2\psi-1\right)}{2(\xi+\eta q^2)}(\zeta+\lambda\gamma\chi(k/k_d))\cos2\psi, \label{eq:Zetaqlim} \\
	\tilde{\chi}_\bq = \chi + \left[\frac{1}{\gamma}+\frac{q^2\left(\lambda\cos 2\psi-1\right)^2}{2(\xi+\eta q^2)}\right]\nu.
\end{gather}

The dynamics of the joint field $\Xi$ and the relative angular separation $\Delta$ can be written as
\begin{gather}
	\partial_t\tilde{\Xi} =-\frac{\tilde{K}_\bq+\tilde{\zeta}_\bq}{2}\tilde{\Xi} -\left( \frac{\tilde{\chi}_\bq k}{k_d}-k_d+\frac{\tilde{K}_\bq +\tilde{\zeta}_\bq}{2}\right)\tilde{\Delta}, \\
	\partial_t\tilde{\Delta} = -\left(\frac{\tilde{\chi}_\bq k}{k_d}+k_d+\frac{\tilde{K}_\bq+\tilde{\zeta}_\bq}{2}\right)\tilde{\Delta}-\frac{\tilde{K}_\bq+\tilde{\zeta}_\bq}{2}\tilde{\Xi}\,.
\end{gather}
The eigenvalues of the associated Jacobian matrix are the growth rates $s_\bq$:
\begin{equation}
	{s_\bq}_\pm=-\frac{1}{2}\left(\frac{\tilde{\chi}_\bq k}{k_d}+k_d+k_d(\tilde{K}_\bq+\tilde{\zeta}_\bq)\pm\sqrt{\left[\tilde{\chi}_\bq k/k_d+k_d+k_d(\tilde{K}_\bq+\tilde{\zeta}_\bq)\right]^2-4k_d(\tilde{K}_\bq+\tilde{\zeta}_\bq)}\right). \label{SI:eigsslab}
\end{equation}
Since $\tilde{K}_\bq$ and $\tilde{\zeta}_\bq$ are both $\mathcal{O}(q^2)$, it is clear that one of the two eigenvalues has non-vanishing relaxation rate in the limit of vanishing wavenumber, i.e., ${s_\bq}_+\sim\mathcal{O}(q^0)$ while the other eigenvalue vanishes at $q=0$: ${s_\bq}_-\sim\mathcal{O}(q^2)$. Indeed, this is apparent from the form of the dynamics of $\tilde{\Xi}$ and $\tilde{\Delta}$: it is evident that $\tilde{\Delta}$ relaxes to $\sim q^2\tilde{\Xi}$ in a finite time. Therefore, when we solve for $\tilde{\Delta}$ and obtain an equation for $\tilde{\Xi}$ alone, its relaxation rate goes as $\sim q^2$ as it should.

When there is no friction, $\xi = 0$, the active term $\tilde{\zeta}_\bq$ is now $\mathcal{O}(q^0)$. In this case it is easy to see that will be a generic instability when $\tilde{\zeta}_\bq < 0$, as is the case in active nematics without a footprint~\cite{Simha2002}. Examining \eqref{eq:Zetaqlim}, we see that we can always find a value of $\psi$ for which $\tilde{\zeta}_\bq<0$, and so the system is \textit{always} unstable.

\subsection{Slab geometry: pure splay instability threshold.}
In the main text examine the linear stability of a uniform state in a quasi-1D slab geometry with strong anchoring boundary conditions ($\theta_0(0)=\theta_0(L)=\pi/2$), where we are searching for instability in the $x$-direction, transverse to the slab's length.
Although Eq.~\eqref{SI:eigsslab} has been calculated for an $x$-aligned state ($\theta_0=0$), we can calculate the threshold for instability by setting $\psi=\pi/2$ and considering perturbations in the transverse $y$-direction, with the boundary conditions $\theta(0)=\theta(L)=0$.
We will obtain the same threshold in both cases.
By examining Eq.~\eqref{SI:eigsslab}, it is clear that for $\tilde{\chi}_\bq k/k_d + k_d > 0$, the growth rate will be positive when $\tilde{K}_\bq+\tilde{\zeta}_\bq < 0$. From this, it follows that in a quasi-1D strip, where $\psi=\pi/2$ is imposed, the instability of the smallest allowed wavenumber $q=\pi/L$ occurs at a threshold width $L_c$ which satisfies 
\begin{equation}
	K\left[\frac{1}{\gamma}+\frac{(\lambda+1)^2}{2(\xi+\eta (\pi/L_c)^2)}\frac{\pi^2}{L_c^2}\right]+\frac{(\lambda+1)(\zeta+\lambda\gamma\chi(k/k_d))}{2(\xi+\eta (\pi/L_c)^2)}=0.
\end{equation}
Multiplying through by $\xi + \eta(\pi/L_c)^2$, and rearranging, we recover the expression for the threshold in the main text:
\begin{equation}
	\left(\frac{\pi}{L_c}\right)^2 = -\frac{\gamma(\lambda+1)(\zeta+\lambda\gamma\chi(k/k_d))}{2K\tilde{\eta}}-\frac{\xi}{\tilde{\eta}}\,,
\end{equation}
where $\tilde{\eta}=\eta+\gamma(\lambda+1)^2/2$.
\subsection{Static matrix}
In the case of static matrix, the stability of the system is given directly from the equation
\begin{equation}
	\partial_t \tilde{\theta} = -\left[\tilde{K}_\bq+\tilde{\zeta}^\mathrm{static}_\bq+\tilde{\chi}_\bq\right]\tilde{\theta} \, ,
\end{equation}
where $\tilde{\zeta}^\mathrm{static}_\bq$ has a modified matrix-induced renormalization of the activity, with no factor of $k/k_d$:
\begin{equation}
	\tilde{\zeta}^\mathrm{static}_\bq = \frac{q^2\left(\lambda\cos 2\psi-1\right)}{2(\xi+\eta q^2)}(\zeta+\lambda\gamma\chi)\cos2\psi\, .
\end{equation}
\subsubsection{Negligible friction}
As in the main text, we consider the case where $\Q$ is deep in the nematic state, such that $S_0=1$.
In the case of zero friction, $\xi = 0$, the usual zero-wavenumber generic instability of active nematic fluids is modified due to static matrix and acquires a threshold in either activity or flow-coupling. The growth rate for the $q=0$ mode is given by
\begin{equation}
	s_{\mathbf{q}= 0}=-\left(\lambda\cos 2\psi-1\right)\frac{(\zeta+\lambda\gamma\chi) \cos 2\psi}{2\eta}-\left(\chi + \left[\frac{1}{\gamma}+\frac{\left(\lambda\cos 2\psi-1\right)^2}{2\eta}\right]\nu\right), \label{SI:eigstatic}
\end{equation}
One interesting limit is the case with no active alignment, $\chi = 0$.
In this case, the threshold for instability is given by
\begin{equation}
	-\left(\lambda\cos 2\psi-1\right)\zeta \cos2\psi>\left[\frac{2\eta+\gamma\left(\lambda\cos 2\psi-1\right)^2}{\gamma}\right]\nu.
\end{equation}
The right hand side of this inequality is always positive, and so coupling to a static matrix in the free energy stabilises the $q=0$ mode: the critical activity for instability is non-zero, contrary to the case without a matrix. Indeed, here $\nu$ acts as an external symmetry breaking field.

As a separate limit, we introduce the matrix coupling only through the active aligning term $\chi$:
\begin{equation}
	s_{\mathbf{q}=0} = -\left(\lambda\cos 2\psi-1\right)\frac{\zeta+\lambda\gamma\chi}{2\eta}\cos2\psi-\chi.
\end{equation}
Interestingly, in this case, $\chi$ may have a \emph{destabilising} effect, though it promotes alignment with a matrix. To see this, let us consider the extreme case with $\zeta=0$. In this case, the condition for an instability is 
\begin{equation}
	[\{(\lambda\cos2\psi)^2-\lambda\cos2\psi\}\gamma+2\eta]\chi<0.
\end{equation}
Taking $\chi>0$ (which implies an alignment with the matrix), this condition is fulfilled for large enough $\gamma/\eta$. To see why this happens physically, notice that $\gamma$ controls the timescale of relaxation of an angular fluctuation to its steady-state value. The ratio of viscosity and active stress is another timescale that controls the growth of fluctuations in this system (it need not always do so; it can instead control the \emph{decay} of fluctuations \cite{AM_interface} or the \emph{time period} of oscillations \cite{AM_col} in other systems). Here, $\chi$ plays the dual role of an active aligning torque and an active destabilising stress. Therefore, whenever $\gamma\gg\eta$, the destabilising aspect of $\chi$ through its involvement in active stress dominates over its stabilising aspect, leading to an instability of the ordered phase.

\subsubsection{Negligible viscosity}

In the limit of zero viscosity, $\eta = 0$, both the only-passive and only-active limits stabilise the system at zero-wavevector relative to the case without matrix:
\begin{equation}
	s_{\bq=0} = -\left(\chi + \nu/\gamma\right)\,.
\end{equation}
In this case, there can only be an intermediate wavelength instability dictated by the magnitude of the effective activity. Since here we deal with a linear hydrodynamic theory, we can not rule out the possibility that non-linear terms will qualitatively modify the higher-order terms in $q$.
\section{Non-reciprocal interactions}
\subsection{Doubly nematic theory}
For nematic order parameters $\Q=S(\hat{\mathbf{n}}\otimes\hat{\mathbf{n}}-\bsf{I}/2)$, $\M=S_M(\hat{\mathbf{m}}\otimes\hat{\mathbf{m}}-\bsf{I}/2)$, where $\hat{\mathbf{n}}=(\cos 2\theta,\sin 2\theta)$, $\hat{\mathbf{m}}=(\cos 2\phi,\sin 2\phi)$, the equations
\begin{gather}
	\partial_t\Q = -\frac{\alpha}{\gamma}\Q-\frac{\beta}{\gamma}\Q^3 + \tilde{\chi}\M \label{Qeqnnrec},\\
	\partial_t \M = k\Q-k_d \M\label{Meqnnrec},
\end{gather}
can be written in terms of $S$, $S_M$, and transformed angular coordinates $\Delta = \theta - \phi$, $\Xi = \theta + \phi$:
\begin{gather}
	\partial_t S = -\frac{\alpha}{\gamma}S - \frac{\beta}{\gamma}S^3 + \tilde{\chi}S_M\cos 2\Delta \label{SI:QMS},\\
	\partial_t S_M  = kS\cos 2\Delta - k_d S_M \label{SI:QMSM},\\
	\partial_t \Delta = \frac{1}{2}\left(-\frac{\tilde{\chi} S_M}{S}-\frac{k S}{S_M}\right)\sin 2\Delta \label{SI:QMpsi},\\
	\partial_t \Xi = \frac{1}{2} \left(\frac{k S}{S_M}-\frac{\tilde{\chi}S_M}{S}\right)\sin 2\Delta. \label{SI:QMvp}
\end{gather}
Eqs.~\eqref{SI:QMS}--\eqref{SI:QMpsi} form a closed system. From Eq.~\eqref{SI:QMpsi}, we see two possibilities for a steady state: either $\Delta=0$, or, when $\tilde{\chi}<-k$, such that $-\tilde{\chi}S_M/S-kS/S_M=0$.

\subsubsection{Aligned state}
When $\Delta=0$, Eq.~\eqref{SI:QMSM} gives that at steady state, $kS=k_dS_M$, and so we immediately find the steady state equation for $S$:
\begin{equation}
	-\left(\frac{\alpha}{\gamma}-\tilde{\chi}(k/k_d)\right)S - \frac{\beta}{\gamma}S^3=0.
\end{equation}
So, this state is described completely by $S=S_M=\sqrt{(\gamma\tilde{\chi}(k/k_d)-\alpha)/\beta}$, and $\Delta=0$.

\subsubsection{Nonaligned state}

We can also satisfy steady state by fulfilling the condition $\tilde{\chi}S_M/S+k S/S_M=0$.
Note that Eq.~\eqref{SI:QMSM} requires that $k S\geq k_d S_M$ in steady state, since $\cos\Delta \leq 1$.
In this case, $S_M/S = \cos 2\Delta (k/k_d)$, $\cos 2\Delta= k_d\sqrt{-1/\tilde{\chi}k}$.
Substituting for $S_M$ and $\cos 2\Delta$ into Eq.~\eqref{SI:QMS}, we find the steady state value for $S$ is then given by
\begin{equation}
	-\left(\frac{\alpha}{\gamma}+k_d \right)S - \frac{\beta}{\gamma}S^3=0.
\end{equation}
Thus, the state is given by $S=\sqrt{-(\alpha+\gamma k_d)/\beta}$, $S_M = S\sqrt{-k/\tilde{\chi}}$ and $\cos 2\Delta = k_d\sqrt{-1/\tilde{\chi}k}$.
Note that, since $k_d>0$, $\alpha$ has to be large and negative.
In contrast to the aligned state, the joint dynamics is not stationary.
While dynamical equations of the form \eqref{Qeqnnrec} and \eqref{Meqnnrec} can also arise in equilibrium materials with distinct dissipative kinetic coefficients for ${\bsf Q}$ and ${\bsf M}$, this possibility doesn't exist. For this state to exist, $k$ and $\tilde{\chi}$ has to have opposite signs which is impossible in equilibrium systems where the Onsager dissipative coefficients must be positive definite.
Substituting the state into Eq.~\eqref{SI:QMvp}, the joint dynamics is given by
\begin{equation}
	\partial_t \Xi = \sqrt{-k\tilde{\chi}}
	\sqrt{1+\frac{k_d^2}{\tilde{\chi}k}}=\sqrt{-(\tilde{\chi}k+k_d^2)}\equiv 2\omega.
\end{equation}
As such, $\Xi=2\omega t$, and the director field angles are given by $\theta=\omega t$, $\phi = \omega t - \Delta$: the two fields rotate with a fixed phase difference. A linear stability analysis about this state gives:
\begin{gather}
	\partial_t \delta S =\left( -\frac{2|\alpha|}{\gamma}+3k_d\right)\delta S+k_d^2\sqrt{-\frac{1}{k\chi}}\delta S_M+\sqrt{\frac{-|\alpha|+k_d\gamma}{\beta}(k_d^2+k\chi)}\delta\Delta\,,\\
	\partial_t\delta S_M=-k_d\delta S_M+k_d\sqrt{-\frac{k}{\chi}}\delta S-2k\sqrt{-\frac{k_d\gamma-|\alpha|}{\beta}\left(1+\frac{k_d^2}{k\chi}\right)}\delta\Delta\,,\\
	\partial_t\delta\Delta=-\sqrt{\frac{\beta(k_d^2+k\chi)}{k^2(k_d\gamma-|\alpha|)}}\left(k\delta S+\sqrt{-k\chi}\delta S_M\right)\,.
\end{gather}
The coupled eigenvalues of this system of linear equations is, in general, complicated. However, in the limit of very large $|\alpha|$, the $\delta S$ fluctuations relax infinitely fast to a value determined by $\delta S_M$ and $\delta\Delta$. The other two eigenvalues go to a non-zero limit for $\alpha\to-\infty$: $(1/2)\left[-k_d\pm\sqrt{9k_d^2+8k\chi}\right]$. Since for the rotating state to exist, $|k\chi|>k_d^2$ and $k\chi<1$, both these eigenvalues clearly have negative real parts. Thus, we have demonstrated that the rotating state is stable at least in the limit of fast relaxation of $\delta S$ fluctuations.

\subsection{Doubly polar theory}
For a completely polar theory $\bp=p(\cos \theta,\sin\theta)$, $\bfm=m(\cos\phi,\sin\phi)$, the equations
\begin{gather}
	\partial_t \bp = -\frac{\alpha}{\gamma}\bp-\frac{\beta}{\gamma}\bp^3 + \tilde{\chi}\bfm , \\
	\partial_t \bfm = k\bp-k_d\bfm,
\end{gather}
may be written in terms of the transformed angular coordinates ($\psi$, $\varphi$) as
\begin{gather}
	\partial_t p = -\frac{\alpha}{\gamma}p - \frac{\beta}{\gamma}p^3 + \tilde{\chi}m\cos\psi \label{SI:pmp} \\
	\partial_t m = kp\cos\psi - k_d m, \label{SI:pmm} \\
	\partial_t \psi =\left(-\frac{kp}{m}-\frac{\tilde{\chi}m}{p}\right)\sin\psi, \label{SI:pmpsi} \\
	\partial_t \varphi = \left(\frac{k p}{m}-\frac{\tilde{\chi}m}{p}\right)\sin\psi.
\end{gather}
The structure of the equations is identical to the $\Q$--$\M$ theory--the only difference is that for given parameters, the phase shift will be twice as large in the $\bp$--$\bfm$ system, and can reach $\psi = \pi/2$.

\subsection{Active polar with nematic tracks theory}
For a polar active field $\bp=p(\cos\theta,\sin\theta)$, with nematic passive field $\M$, the equations are written
\begin{gather}
	\partial_t \bp = -\frac{\alpha}{\gamma}\bp-\frac{\beta}{\gamma}\bp^3 + \tilde{\chi}\M\bp,\\
	\partial_t \M = k\tilde{\bsf{P}}-k_d\M,
\end{gather}
where $\tilde{\bsf{P}}=\bp\otimes\bp - |p|^2\bsf{I}/2$. Note that here we neglect the $\bp^5$ term since we are considering $\tilde{\chi}<0$.
As above, we use the transformed angular variables ($\psi$, $\varphi$) to rewrite these equations in $p$ and $S_M$:
\begin{gather}
	\partial_t p= -\frac{\alpha}{\gamma}p-\frac{\beta}{\gamma}p^3 + \tilde{\chi}S_M p \cos 2\psi, \label{SI:pMp} \\
	\partial_t S_M = k p^2\cos 2\psi - k_d S_M, \label{SI:pMSM} \\
	\partial_t \psi = \left(-\frac{kp^2}{2S_M} - \tilde{\chi}S_M\right)\sin 2\psi, \label{SI:pMpsi} \\
	\partial_t \varphi = \left(\frac{k p^2}{2S_M}-\tilde{\chi}S_M\right)\sin 2\psi. \label{SI:pMvp}
\end{gather}
An aligned state, $\psi=0$, has order parameter $p=\sqrt{-\alpha/(\beta-\gamma\tilde{\chi}k/k_d)}$.
We consider the case of nonreciprocal interactions ($\tilde{\chi}<0$), where this state is only stable when $\alpha < 0$.
Given that $S_M=k p^2/k_d$ in the aligned state, Eq.~\eqref{SI:pMpsi} gives the condition for instability of the angular variable in the aligned state:
\begin{equation}
	-\frac{k_d}{2} + \frac{\tilde{\chi}\alpha k/k_d}{\beta-\gamma\tilde{\chi}k/k_d}>0,
\end{equation}
or, written explicitly as a condition on $\tilde{\chi}$:
\begin{equation}
	\tilde{\chi} < \frac{k_d^2\beta}{k(2\alpha + \gamma k_d)}.
	\label{SI:pMchi}
\end{equation}
However, we can also search for other steady state solutions, using $\cos 2\psi=k_d S_M/k p^2$, and $-kp^2/(2S_M) - \tilde{\chi}S_M=0$. In this case, Eq.~\eqref{SI:pMp} gives the expression for the steady state value of $p$:
\begin{equation}
	-\left(\frac{\alpha}{\gamma}+\frac{k_d}{2}\right)p - \frac{\beta}{\gamma}p^3 = 0.
\end{equation}
Therefore, $p=\sqrt{-(\alpha+\gamma k_d/2)/\beta}$.
The condition for the appearance of this state is that $\cos 2\psi < 1$, i.e.:
\begin{equation}
	\frac{k_d}{k}\sqrt{\frac{\beta}{-(\alpha+\gamma k_d/2)}}\sqrt{\frac{k}{-2\tilde{\chi}}} < 1.
\end{equation}
Squaring both sides and rearranging:
\begin{equation}
	\tilde{\chi} < \frac{k_d^2\beta}{k(2\alpha+ \gamma k_d)}.
\end{equation}
This is identical to the condition for instability of the aligned state in Eq.~\eqref{SI:pMchi}.
In other words, when the aligned state becomes unstable, a new state with a finite phase shift $\psi$ between the director field directions establishes itself.
As in the $\Q$--$\M$ theory, this is a rotating state. Note that it is not possible to transition directly to the rotating state from the isotropic state!

\subsection{Oscillating shear in a slab geometry: Hopf bifurcation analysis}

As already detailed in Eq.~\eqref{SI:eigsslab}, upon performing a linear stability analysis about a uniform state in a slab geometry with strong anchoring boundary conditions ($\theta_0(0)=\theta_0(L)=\pi/2$), the eigenvalues of the associated Jacobian matrix are the growth rates $s_\bq$:
\begin{equation}
	{s_\bq}_\pm=-\frac{1}{2}\left(\frac{\tilde{\chi}_\bq k}{k_d}+k_d+k_d(\tilde{K}_\bq+\tilde{\zeta}_\bq)\pm\sqrt{\left[\tilde{\chi}_\bq k/k_d+k_d+k_d(\tilde{K}_\bq+\tilde{\zeta}_\bq)\right]^2-4k_d(\tilde{K}_\bq+\tilde{\zeta}_\bq)}\right).
\end{equation}
We consider a regime where $\tilde{K}_\bq+\tilde{\zeta}_\bq>0$, but now allow the footprint coupling $\chi$ (and consequently $\tilde{\chi}_\bq$) to be negative. There is an instability when
\begin{equation}
	\tilde{\chi}_\bq < -\frac{k_d^2}{k}\left(1+\tilde{K}_\bq+\tilde{\zeta}_\bq\right)\,.
\end{equation}
At the transition point, the eigenvalues are purely imaginary:
\begin{equation}
	s_{\bq_\pm} = \pm i\sqrt{k_d(\tilde{K}_\bq+\tilde{\zeta}_\bq)}\,,
\end{equation}
indicating that the transition is a Hopf bifurcation.

\section{Numerical study}

The numerical study was performed using the Finite Element Method software Comsol Multiphysics. 
We wrote weak forms of the equations in the software physics \texttt{Weak Form PDE}.\\

In Fig.~3 and Fig.~4, we set no-slip boundary conditions, together with an anchoring $\theta=\pi/2$ to the boundaries.
The values of the parameters for Fig.~3-4 are shown in Table.~\ref{tab:supFig}.

\begin{table}[h!]
	\label{tab:supFig}    \centering
	\begin{tabular}{|c|c|c|c|}
		\hline
		& Fig.~3b & Fig.~3c/Supp. Movie 1 & Fig.~4/Supp. Movie 2  \\ 
		\hline
		$K$ & $0.1$ & $0.1$ & $0.1$ \\
		$\alpha$& $-100$ &$-100$ & $-1$ \\
		$\beta$& $100$  & $100$ & $1$ \\
		$\chi$& -- & $2$ & $0$\\
		$\nu$& $0$ & $0$ & $-0.1$\\
		$\gamma$& $1$ & $1$ & $0.05$ \\
		$\lambda$& -- & $2$ & $0$ \\
		$\eta$& $1$ & $1$ & $100$ \\
		$\zeta$& -- &$-1$ & $150$ \\
		$\xi$& $0$ &$0$ & $0$\\
		$k$& $1$ & $0.01$ & $1$ \\
		$k_d$& $1$ & $0.01$ & $1$ \\
		$L $& $3.14$ & $6.6$ & $10.5$ \\
		\hline
	\end{tabular}
	\caption{Parameters used in the numerical studies shown in Fig.~3-4, and the accompanying Supplementary Movies.}
	\label{tab:param}
\end{table}

\vspace{1em}

\noindent\textbf{Supplementary Movie 1:} A movie representation of Fig.~3c, with the same parameters as in the main text, given in Table~\ref{tab:param}.

\vspace{1em}

\noindent\textbf{Supplementary Movie 2:} A movie representation of Fig.~4a, with the same parameters as in the main text, given in Table~\ref{tab:param}.

\bibliography{refs}

\end{document}